\documentclass[aps,preprint,nofootinbib,superscriptaddress]{revtex4-1}

\usepackage{graphicx}  %
\usepackage{amsmath}
\usepackage{amsfonts}
\usepackage{amssymb}
\usepackage{slashed}
\usepackage{ulem}
\usepackage{bm}  %
\usepackage{color}
\usepackage{xcolor}
\usepackage{hyperref}
\hypersetup{
    colorlinks,
    linkcolor={red!50!black},
    citecolor={blue!50!black},
    urlcolor={blue!80!black}
}

\newcommand{\bea}{\begin{eqnarray}}
\newcommand{\eea}{\end{eqnarray}}
\newcommand{\beq}{\begin{equation}}
\newcommand{\eeq}{\end{equation}}
\newcommand{\be}{\begin{equation}}
\newcommand{\ee}{\end{equation}}
\newcommand{\bqa}{\begin{eqnarray}}
\newcommand{\eqa}{\end{eqnarray}}

\makeatletter
\newcommand{\lrhup}[2]{%
  \ooalign{$\m@th#1\leftharpoonup$\cr$\m@th#1\rightharpoonup$\cr}%
}
\makeatother
\def\mqo2{{\!\!\!}}

\newcommand{\cM}{\mathcal{M}}
\newcommand{\cI}{\mathcal{I}}
\newcommand{\cJ}{\mathcal{J}}

\newcommand{\cZd}{\mathcal{Z}_{d}}

\newcommand{\ALO}{\mathcal{A}_{\rm LO}}
\newcommand{\ANLO}{\mathcal{A}_{\rm NLO}}
\newcommand{\ANNLO}{\mathcal{A}_{\rm NNLO}}

\newcommand{\eq}[1]{Eq.~\eqref{eq:#1}}

\newcommand{\nn}{\nonumber}

\newcommand{\larrowpartial}[1]{\overleftarrow{\partial}\!\!_{#1}}
\newcommand{\rarrowpartial}[1]{\overrightarrow{\partial}\!\!_{#1}}

\newcommand{\bs}[1]{\boldsymbol{#1}}

\begin{document}
\title{Pionless Effective Field Theory Evaluation of Nuclear Polarizability in Muonic Deuterium}
\author{Samuel B. Emmons}
\affiliation{Department of Mathematics, Physics, and Computer Science,
Carson-Newman University, Jefferson City, TN 37760, USA}
\affiliation{Department of Physics and Astronomy, University of
Tennessee, Knoxville, TN 37996, USA}
\author{Chen Ji}
\email{jichen@mail.ccnu.edu.cn}
\affiliation{Key Laboratory of Quark and Lepton Physics, Institute of Particle Physics, Central China Normal University, Wuhan 430079, China}
\author{Lucas Platter}
\affiliation{Department of Physics and Astronomy, University of
Tennessee, Knoxville, TN 37996, USA}
\affiliation{Physics Division, Oak Ridge National Laboratory, Oak Ridge, TN 37831, USA}
\date{\today}

\begin{abstract}
  We calculate the longitudinal structure function
  of the deuteron up through next-to-next-to-leading order in the
  framework of pionless effective field theory.
  We use these results to compute the two-photon polarizability
  contribution to Lamb shift in muonic deuterium, which can be 
  utilized to extract the nuclear charge radius of the deuteron. 
  We present analytical expressions order-by-order for the relevant transition 
  matrix elements and the longitudinal structure function,
  and we give numerical results for the corresponding contributions to 
  the Lamb shift. We also discuss the impact of relativistic and other 
  higher-order effects. We find agreement with previous
  calculations and explain the accuracy of our calculation. 
\end{abstract}

\smallskip
\maketitle

\newpage
\section{Introduction}
The deuteron is the simplest bound nuclear system and is made up of
only two nucleons. It is the perfect testing ground for new ideas in
nuclear theory since calculations are relatively simple for this
system and can be compared with decades of experimental data. The
electromagnetic properties of a nucleus provide insights into
its size, shape, and its continuum properties. These are
therefore sensitive observables that test our understanding of short-
and long-range charactistics of the nucleon-nucleon interaction.

The charge radius is one of the most elementary electromagnetic
observables. It can be measured using either elastic electron-nucleus
scattering or laser spectroscopy. An analysis of the world-averaged
electron-deuteron scattering data determined the deuteron
root-mean-square (rms) charge radius to be
$r_d = 2.130(10)$~fm~\cite{Sick2008}. The transition frequencies among
atomic levels in deuterium depend on $r_d$ and an analysis based on
the world-averaged deuteron spectroscopy data yields
$r_d = 2.1415(45)$~fm~\cite{Pohl:2016glp}. A recent measurement of the
Lamb shift in muonic deuterium ($\mu$-d) found a smaller radius,
$r_d =2.12562(78)$~\cite{Pohl2016}, which deviates by $3.5\sigma$ from
the electronic deuterium ($e$-d) spectroscopic result. 
This apparent difference of $r_d$ in $e$-d and
$\mu$-d coincides with the original proton radius puzzle
that spurred much theoretical and experimental work after
a $7\sigma$ deviation was discovered between the proton charge 
radius extracted from the $\mu$-H Lamb shift,
$r_p=0.84087(39)$~fm~\cite{Pohl2010,Antognini2013}, and 2016 
world-averaged hydrogen spectroscopy data giving $r_p=0.8759(77)$
fm~\cite{Mohr2016}. Three new spectroscopy experiments have been
conducted using electronic hydrogen, two of 
which~\cite{Beyer:2017gug,Bezginov:2019mdi}
agree with the smaller proton radius and one of 
which~\cite{Fleurbaey:2018fih} agrees with the the larger value. 
Furthermore, recent electron-proton scattering data from
the PRad experiment~\cite{Xiong:2019umf} suggest a smaller proton
radius. These new experiments point to a possible resolution of the
radius puzzle. However, explanation of discrepancies among different
experiments is still needed.

The determination of the nuclear charge radius from $\mu$-d spectroscopy 
is sensitive to the two-photon exchange (TPE) contribution to
the atomic $2S$-$2P$ level spacing~\cite{Borie2012}. 
The electromagnetic polarization of the nucleus
caused by the muon leads to a distortion of the muon-nucleus wave
function and affects thereby the atomic spectrum. In the TPE
process, the nucleus is virtually excited and de-excited by
the exchange of two photons with the muon. Therefore, TPE depends not
only on the bound-state properties of the deuteron, but also on the
nucleon-nucleon continuum scattering state and the form of the
electromagnetic current. TPE plays a crucial role in connecting the
physics of nuclear structure to photonuclear reactions, 
and the accuracy of related calculations depends how well
known the nuclear Hamiltonian is.

The effects of TPE on $\mu$-d observables were originally calculated
in Refs.~\cite{Pachucki1993,Rosenfelder1983,Lu1993,Leidemann1995} and
were recently revisited with improved accuracy using different nuclear
models~\cite{Pachucki2011,Pachucki2015,Hernandez2014,Hernandez2018,
  Hernandez:2019zcm,Friar2013}.  The calculations done with Argonne
V18, chiral effective field theory ($\chi$EFT), and zero-range
approximated (ZRA) nucleon-nucleon interactions show good agreement
with each other, demonstrating the high predictive power and accuracy
of the state-of-the-art nuclear models.  By analyzing statistical and
systematic uncertainties on TPE calculations performed within the
$\chi$EFT framework, the uncertainty due to nuclear model
  dependence in these calculations was
probed~\cite{Hernandez2014,Hernandez2018}. Furthermore, TPE in muonic
deuterium was also considered using a dispersion relation analysis of
the scattering data~\cite{Carlson2014}. This approach was also shown
to agree well with the aforementioned nuclear model calculations.
Additional work has extended the evaluation of TPE to other light
muonic atoms and ions, {\it i.e.}, $\mu^3{\rm H}$, $\mu^3{\rm He}^+$,
and
$\mu^4{\rm
  He}^+$~\cite{Friar1977,Ji2013,Dinur2016,Carlson2017,Ji:2018ozm}.

In this work, we make use of the work of Rosenfelder and
Leidemann~\cite{Rosenfelder1983,Leidemann1995} and thereby also more
recent calculations that use a state-of-the-art nuclear
Hamiltonian~\cite{Pachucki2011,Hernandez2014,Hernandez:2019zcm}. However,
instead of $\chi$EFT, we will use pionless effective field theory
($\slashed{\pi}$EFT). Calculations in this framework can be expanded
on an order-by-order basis in an expansion parameter that is
proportional to the range of the nuclear interaction $R$ over the
two-nucleon scattering length $a$.  The momentum scale of the
processes considered within this approach are assumed to be of order
$1/a$. $\slashed{\pi}$EFT is suited for processes whose momentum
scales are well below the pion mass, as in $\mu$-d.  For such
processes, this approach offers a systematic expansion that is order
by order renormalizable, results whose regulator dependence is
transparent and understood, and for the two-nucleon systems frequently
also analytic results that reveal the dependence on physical
parameters directly~\cite{Hammer:2019poc}.  Additionally, since the
muon is approximately 200 times heavier than the electron, it orbits
closer to the nucleus and may be considered as approximately
non-relativistic, enabling us to neglect relativistic effects at the
EFT order we consider.  Compared with $\chi$EFT, the order-by-order
renormalizability and regulator independence in $\slashed{\pi}$EFT
provides a rigorous systematic uncertainty estimation which is
model-independent. Further, the smaller number of parameters in
$\slashed{\pi}$EFT make it a powerful tool to explore few-body
universality in few-nucleon systems \cite{Hammer2010a}.

The organization of this work is as follows. We introduce the basic
equations that relate the inelastic structure functions and electric
form factors to the TPE effect in muonic deuterium in 
Sec.~\ref{sec:tpe-th}. In Sec.~\ref{sec:pionl-effect-field}, 
we explain the pionless EFT Lagrangian that will be utilized in 
this work. We then show in Sec.~\ref{sec:diagr-calc-struct} how
diagrammatic calculations can be used to obtain the inelastic
structure function. In Sec.~\ref{sec:Results}, we present results
for the TPE energy shift obtained from our calculations and compare it to
previous calculations. We conclude with a summary and a discussion of 
possible extensions of this work.

\section{Theory of two-photon exchange contributions}
\label{sec:tpe-th}
The two-photon exchange contributions in muonic deuterium can be separated 
into a part depending on the structure of the atomic nucleus and another 
that depends on the internal dynamics of the single nucleon. In this paper, 
we focus only on the nuclear two-photon exchange contribution to the 
former, labeled $\delta_{\rm TPE}^A$, by considering single nucleons as 
point-like particles. $\delta_{\rm TPE}^A$ consists of the elastic and the 
inelastic parts
\begin{equation}
\label{eq:deltaTPE}
\delta_{\rm TPE}^{A} = \delta_{\rm Zem}^{A} + \delta_{\rm pol}^{A}~,
\end{equation}
where the elastic part $\delta^A_{\rm Zem}$ corresponds to the nuclear
third Zemach moment contribution 
first derived for light muonic atoms by Friar~\cite{Friar1979} as a
nuclear finite-size contribution of order $\alpha^5$, where
$\alpha=1/137.036$ denotes the fine structure constant.  It is given
by~\cite{Sick2014,Nevo2019}
\begin{align}
\label{eq:Zem}
\delta^{A}_{\rm Zem} =
-m_{r}^4\frac{\alpha^5}{24}\left\langle R_{\text{E}}^{3}\right\rangle_{(2)}=&
-m_{r}^4\alpha^5\frac{2}{\pi}\int_{0}^{\infty}
\frac{dq}{q^4}
\left[F_{\text{E}}^{2}(q^2)-1+q^2\frac{\langle R_{\text{E}}^2\rangle}{3}\right]~,
\end{align}
where $m_r$ is the muon-deuteron reduced mass,
$F_{\text{E}}(q^2)$ is the deuteron electric form factor, and
$\langle R_{\text{E}}^2\rangle=-6 [\partial F_{\text{E}}(q^2)/\partial q^2]_{q=0}$,
where the derivative is taken with respect to a low-$q^2$ expansion 
of $F_{\text{E}}(q^2)$. 

The inelastic contribution $\delta_{\rm pol}^{A}$ in \eq{deltaTPE}
is due to the electric polarization of the nucleus in which
the deuteron is virtually excited by exchanging two photons with the
muon. This is related to the integral of the forward virtual Compton 
amplitude that can be written in terms of the nuclear inelastic
structure functions \cite{Rosenfelder1983}. The polarizability term
$\delta_{\rm pol}^{A}$ may be further separated into
longitudinal and transverse parts as
\cite{Rosenfelder1983,Leidemann1995}
\begin{equation}
\delta_{\rm pol}^{A} = \delta_{{\rm pol},L}^{A}+\delta_{{\rm pol},T}^{A}~, 
\end{equation}
where $\delta_{{\rm pol},L}^{A}$ and $\delta_{{\rm pol},T}^{A}$ are 
defined respectively as \cite{Leidemann1995}
\begin{eqnarray}
\label{eq:polL}
\delta_{{\rm pol},L}^{A} &=&
-8 \alpha^{2} \vert\phi(0)\vert^{2}
\int_{0}^{\infty} d q \int_{\omega_{th}}^{\infty} d \omega
K_{L}(\omega, q) S_{L}(\omega, q)~,\\
\label{eq:polT}
\delta_{{\rm pol},T}^{A} &=&
-8 \alpha^{2} \vert\phi(0)\vert^{2}
\int_{0}^{\infty} d q \int_{\omega_{th}}^{\infty} d \omega
\left[K_{T}(\omega, q) S_{T}(\omega, q)
+K_{S}(\omega, q) S_{T}(\omega,0)\right]~.
\end{eqnarray}
The variables $(\omega,q)$ are the four-momentum 
carried by the exchanged photon, where $q=|\bs{q}|$, 
$\phi(0)=\sqrt{\alpha^3 m_{r}^3/8\pi}$ is the atomic $2S$-state wave 
function at origin. To ensure that only the inelastic
regime is considered, we have $\omega_{th}\geq B_d + q^2/4m_N$. 
$B_d = 2.2246$~MeV is the deuteron binding energy, $m_N = 938.92$ MeV is 
two times the proton-neutron reduced mass, and $q^2/4m_N$ is the recoil energy 
of the nucleus. $S_{L}$ and $S_{T}$ are the longitudinal
and transverse deuteron inelastic structure functions, respectively.

The longitudinal integration kernel in Eq.~\eqref{eq:polL} is given by
\begin{equation}
\label{eq:KL}
K_{L}(\omega, q)=\frac{1}{2 E_{q}}\left[\frac{1}{\left(E_{q}-m_\mu\right)\left(\omega+E_{q}-m_\mu\right)}-\frac{1}{\left(E_{q}+m_\mu\right)\left(\omega+E_{q}+m_\mu\right)}\right]~,
\end{equation}
where $m_\mu$ denotes the muon mass, and $E_{q} = \sqrt{q^2+m_\mu^2}$
retains the relativistic kinematics of the muon. The 
transverse and seagull kernels of Eq.~\eqref{eq:polT}
are provided in Ref.~\cite{Leidemann1995}.
The \textit{seagull} term is required to ensure the gauge invariance and 
to cancel the infrared singularity near $q=0$ in the transverse term. In 
the Coulomb gauge utilized in this paper, the seagull term contributes 
only to $\delta_{{\rm pol},T}^A$~\cite{Rosenfelder1983}.  In the 
non-relativistic limit $q\ll m_\mu$, the longitudinal kernel in 
Eq.~\eqref{eq:KL} is approximated in $q/m_{\mu}$ expansion by
\begin{equation}
\label{eq:KLNR}
K_{L}^{NR} = \frac{1}{q^2(\omega+q^2/2m_\mu)}~.
\end{equation}
This kernel's higher-order terms 
emerge as relativistic corrections.

We note that the expressions for $\delta_{{\rm pol},L}^{A}$ and
$\delta_{{\rm pol},T}^{A}$ in Eqs.~\eqref{eq:polL} and \eqref{eq:polT}
were multiplied by an additional factor $R^{(\mu)}=0.9778$ in
Ref.~\cite{Leidemann1995} to take into account the modification of
muonic deuterium wave function due to the nuclear finite-size
correction. However, such a correction is formally an $\alpha^6$
effect, and is thus neglected in this paper for consistency since the
evaluation of $\delta_{{\rm pol}}^{A}$ and $\delta_{{\rm Zem}}^{A}$ is
of order $\alpha^5$.

In this work, we compute only relevant contributions up through
next-to-next-to-leading order in $\slashed{\pi}$EFT.  
Following Rosenfelder~\cite{Rosenfelder1983}, we relate the
longitudinal part of the structure function $S_L$ to the transition
matrix element  $\mathcal{M}$ by
\begin{equation}
  \label{eq:SL-def}
  S_L(\omega, \bs{q}) = \int \frac{\hbox{d}^3 p}{(2\pi)^3}
  \delta(\omega-B_d-\frac{q^2}{4m_N} - \frac{p^2}{m_N} )\, 
  \overline{|\cM|^2}~,
\end{equation}
where $\overline{|\cM|^2}$ is the squared transition matrix element
for the electric density operator, between the deuteron ground state
and all intermediate excited states.

In the framework of pionless effective field theory
$\slashed{\pi}$EFT, we will calculate the squared matrix element
$\overline{|\cM|^2}$ relevant to the longitudinal deuteron structure
function by considering the coupling of a single $A_0$ Coulomb photon
to the deuteron. We will discuss below that contributions arising from
the transverse structure function do not contribute to the order
considered here. The deuteron charge form factor $F_E(q^2)$ has been
evaluated using $\slashed{\pi}$EFT in Ref.~\cite{Chen:1999tn}.

\section{pionless effective field theory}
\label{sec:pionl-effect-field}
In this section, we provide a brief overview of $\slashed{\pi}$EFT and the 
partial-divergence subtraction (PDS) renormalization scheme. We 
include some details about both the on-shell and off-shell 
nucleon-nucleon scattering amplitudes that are needed in the 
transition matrix element calculations required for
the structure function calculation of Eq.~\eqref{eq:SL-def}.
\subsection{Lagrangian and Feynman rules}
The nucleonic part of the $\slashed{\pi}$EFT Lagrangian is given by~\cite{Chen:1999tn}
\begin{eqnarray}
  \label{eq:Lagrangian0-NN}
  \mathcal{L}_0 &=& N^\dagger \left (i \partial_0 + \frac{\bs{\nabla}^2}{2 m_N} \right) N
  - C_0 \left(N^T P_i N\right)^\dagger \left(N^T P_i N\right)\nn\\
    &&+\frac{C_2 }{8}\left[
    (N^T P_i N)^\dagger \left(N^T P_i\overleftrightarrow{\nabla}^2 N \right) +\hbox{h.c}
  \right]
  -\frac{C_4}{16} \left( N^T P_i \overleftrightarrow{\nabla}^2 N\right)^\dagger
  \left( N^T P_i \overleftrightarrow{\nabla}^2 N\right)\,.
\end{eqnarray}
where we included the EFT nucleon-nucleon contact interactions in the
$^3$S$_1$-channel up to next-to-next-to-leading order
(NNLO) and $P_i = \sigma_2 \sigma_i\otimes \tau_2/\sqrt{8}$ is the 
spin-isospin projection for the $^3$S$_1$ channel.
Additionally, $P_{i}\overleftrightarrow{\nabla}^2=
P_{i}\rarrowpartial{}^{\,\,2}+
\larrowpartial{}^{\,\,2}P_{i}-2\larrowpartial{}P_{i}
\rarrowpartial{}$~, and
the low-energy constants (LECs) $C_i$ are determined through
renormalization by reproducing parameters in the effective range
expansion around the deuteron pole. The neutron-proton
spin-triplet scattering phase shift is expanded  as
\begin{equation}
\label{eq:phase_shift}
k \cot \delta_t = -\gamma + \frac{\rho_d}{2}(k^2+\gamma^2) + \cdots, 
\end{equation}
where $\gamma = \sqrt{m_N B_d}$ denotes the deuteron binding momentum
and $\rho_{d}= 1.764$~fm is the effective range. Assuming that the
momentum scale of processes considered here is comparable to the
deuteron binding momentum, the expansion parameter in $\slashed{\pi}$EFT is
$\gamma \rho_d \approx 0.4$. We count the nucleon mass as
$m_N \approx (\gamma \rho_d)^{-1} \rho_d^{-1}$.
Feynman rules corresponding to the Lagrangian given in
Eq.~\eqref{eq:Lagrangian0-NN} give a nucleon propagator as
$S(p_0, \bs{p}) = [p_0 -{\bs{p}^2}/{2 m_N}+ i \varepsilon]^{-1}$.  The
first relativistic correction to a two-nucleon matrix element is the
kinetic energy multiplied with a term proportional to
$k^2/m_N^2$. Using $k\sim\gamma$ and the aforementioned counting for
the nucleon mass we see that this correction enters at
$\mathcal{O}\left((\gamma \rho_d)^4\right)$. 

Under the $\slashed{\pi}$EFT expansion, the LECs are expanded analogously by
\begin{eqnarray}
\label{eq:Cexpand}
C_0 &=& C_{0,-1} + C_{0,0} + C_{0,1} \nn~,\\
C_2 &=& C_{2,-2} + C_{2,-1} \nn~,\\
C_4 &=& C_{4,-3}~,
\end{eqnarray}
where for a coefficient $C_{n,m}$, $n$ denotes the power of momentum
in the contact term and $n+m+1$ indicates the $\slashed{\pi}$EFT order at which
$C_{n,m}$ emerges. In the power-divergence-subtraction (PDS)
scheme, the expanded LECs of Eq.~\eqref{eq:Cexpand} are given
by~\cite{Chen:1999tn,Kaplan:1998tg}
\begin{align}
\label{eq:Ci}
C_{0,-1}=&-\frac{4\pi}{m_N}\frac{1}{(\mu-\gamma)}~,
&
C_{2,-2}=&\frac{2\pi}{m_N}\frac{\rho_{d}}{(\mu-\gamma)^2}~,
\nn\\
C_{0,0}=&\frac{2\pi}{m_N}\frac{\rho_{d}\gamma^2}{(\mu-\gamma)^2}~,
&
C_{2,-1}=&-\frac{2\pi}{m_{N}}\frac{\rho_{d}^{2}\gamma^{2}}{(\mu-\gamma)^{3}}~,
\nn\\
C_{0,1}=&-\frac{\pi}{m_{N}}\frac{\rho_{d}^2\gamma^4}{(\mu-\gamma)^{3}}~,
&
C_{4,-3}=&-\frac{\pi}{m_{N}}\frac{\rho_{d}^{2}}{(\mu-\gamma)^{3}}~,
\end{align}
where $\mu$ is the PDS renormalization scale.

The electromagnetic interaction with the nucleon field is included by
replacing the four-gradient with the minimally-coupled gauge covariant
derivative $D_\mu = \partial_\mu + i e \mathcal{Q} A_\mu $, where
$\mathcal{Q} = (1+\tau_{3})/2$ denotes the nucleon charge operator
with Pauli matrix $\tau_{3}$ acting in isospin space and
$A_\mu$ is the electromagnetic
gauge field. The Lagrangian for the induced electromagnetic
interaction is thus given by
\begin{align}
  \label{eq:LagStruct}
  \mathcal{L}_{\rm EM}
  =&-e N^{\dagger} \mathcal{Q} N A_{0}
     +\frac{ie}{2 m_N}\left[N^{\dagger}
     \mathcal{Q}(-\overleftarrow{\nabla}+\overrightarrow{\nabla})N\right] \cdot \bs{A}
     -\frac{e^{2}}{2m_N}N^{\dagger}Q \bs{A}\cdot \bs{A} Q N \,,
\end{align}
where the last term is a two-photon coupling and yields the seagull
term in the two-photon exchange, which contributes only to
$\delta_{{\rm pol},T}$ when Coulomb gauge is
adopted~\cite{Rosenfelder1983}. We also drop the two-nucleon current
from the minimal coupling within the contact interactions, because it
only enters at higher orders than the NNLO considered in this paper.

The nucleon current from one-photon coupling is given by
\begin{equation}
J^\mu (\bs{p},\bs{p}') = e\mathcal{Q} \left(1,\frac{\bs{p}+\bs{p}'}{2m_N}\right).
\end{equation}
We see that the transverse current enters the
transition matrix element with a factor of $k/m_N$. This implies that the
squared matrix element entering the calculation of the transverse
structure factor is also $\mathcal{O}((\gamma\rho_d)^4)$.

\subsection{Leading order nucleon-nucleon amplitude}
\begin{figure}[t]
\begin{center}
\includegraphics[width=0.8\textwidth,clip=true]{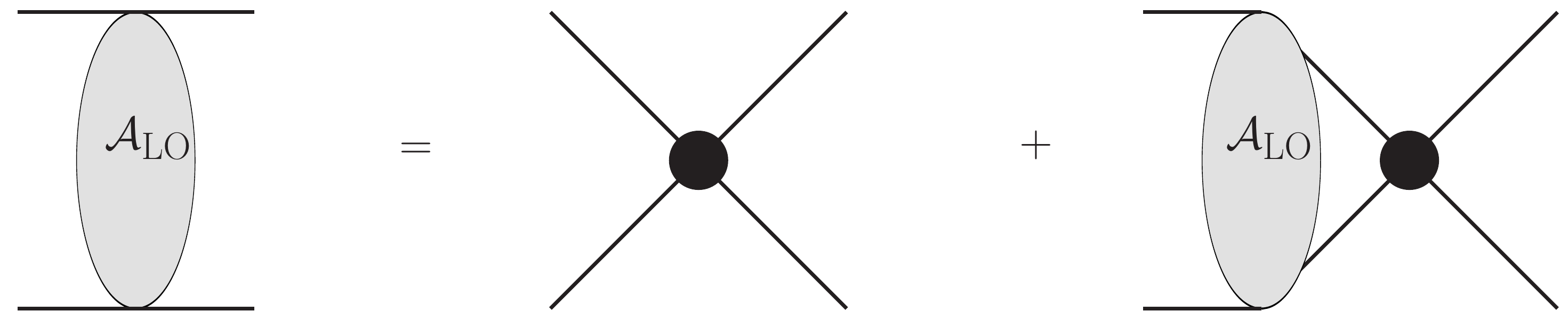}
\end{center}
\caption{\label{fig:ALO} Diagrammatic representation of the
  Lippmann-Schwinger equation for the LO two-body scattering amplitude
  $\ALO$. The round vertex represents insertion of the LO contact
  term.}
\end{figure}

We denote the leading order amplitude in the triplet channel as
$\ALO$, which is shown diagrammatically in Fig.~\ref{fig:ALO}. This is
the Lippmann-Schwinger equation and is solved by identifying the
expression in the figure as the iterative sum
\begin{align}
  \label{eq:ALO-1}
  i \ALO (E) =& -i C_{0,-1}\left[1-\cI_{0}(E)\ALO(E) \right]~,
\end{align}
where $E$ is the two-nucleon energy in the center of mass frame and
$\cI_0$ indicates the loop integral defined in Eq.~\eqref{eq:I2nPDS}.
$\cI_0$ is solved using the PDS scheme \cite{Kaplan:1998tg} and its
dependence on the renormalization scale $\mu$ is given in
Eq.~\eqref{eq:I2nPDS}.  The leading-order amplitude
\begin{equation}
  \label{eq:ALO}
  \ALO (E) = -\frac{4\pi}{m_N} \frac{1}{\gamma + i p} ~,
\end{equation}
is obtained using Eq.~\eqref{eq:ALO-1} and taking the expression of
$C_{0,-1}$ in Eq.~\eqref{eq:Ci}, where $p=\sqrt{m_N E}$ is the 
nucleon-nucleon on-shell relative momentum.
  
\begin{figure}[t]
	\begin{center}
		\includegraphics[width=0.95\textwidth,clip=true]{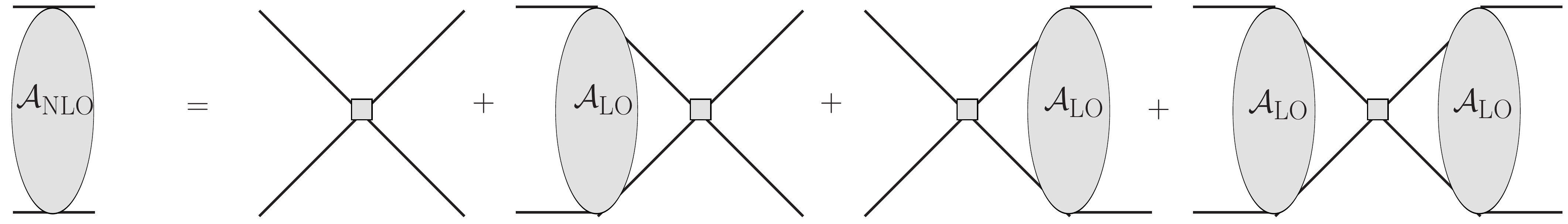}
	\end{center}
	\caption{\label{fig:ANLO} Diagrammatic representation of the
		NLO amplitude calculated perturbatively. The square vertex 
	indicates the insertion of an NLO vertex rule.}
\end{figure}

\subsection{Next-to-leading order amplitude}
The diagrams required to evaluate the next-to-leading-order (NLO) are 
shown in Fig.~\ref{fig:ANLO} and lead to the NLO half-off-shell amplitude 
\begin{align}
  \label{eq:ANLO1}
  i \ANLO (k,p;E) =& - i C_{0,0}\left[ 1+ i \cI_0 i \ALO\right]^2
  -i C_{2,-2} (1+ i\cI_0 i\ALO) \left[\frac{k^2+p^2}{2} + i \cI_2 i\ALO\right]\nn\\
  =& -i \frac{2\pi}{m_N} \frac{\rho_d}{\gamma+i p}
  \left[\gamma-i p+\frac{1}{2(\gamma-\mu)}\left(k^2-p^2\right)\right]~.
\end{align}
where the loop integrals $\cI_0$, $\cI_2$ and the amplitude $\ALO$ are evaluated at the center-of-mass 
energy $E=p^2/m_N$, and $k$ is the incoming-state momentum.
We arrive at the second line of Eq.~\eqref{eq:ANLO1} 
using the Lippmann-Schwinger equation~\eqref{eq:ALO-1}
and using $\cI_{2n} = p^{2n} \cI_0$ from Eq.~\eqref{eq:I2nPDS}.
On the energy shell, the incoming
momentum equals the on-shell momenta $k=p$. This yields the on-shell
NLO amplitude
\begin{equation}
\label{eq:ANLO-onshell}
\ANLO(p,p,E) = -\frac{2\pi \rho_d }{m_N} \frac{\gamma-i p}{\gamma+i p}~.
\end{equation}

\subsection{Next-to-next-to-leading order amplitude}
To obtain the NNLO amplitude, we make use of a diagrammatic expression
including all NNLO diagrams. After replacing the couplings 
in the resulting expression using 
Eq.~\eqref{eq:Ci}, we find the half-off-shell NNLO amplitude to be
\begin{equation}
\label{eq:NNLOhalf}
\ANNLO(k, p; E)
=-\frac{\pi }{m_N} \frac{\rho_d^{2}}{\gamma+i p}
\left[(\gamma-i p)^2  + \frac{\gamma-i p}{\gamma-\mu}  \left(1+\frac{\gamma+i p}{\gamma-\mu}\right)
\frac{k^{2}-p^{2}}{2}\right]~.
\end{equation}
The NNLO on-shell amplitude is found after setting $k=p$ and is
\begin{equation}
\label{eq:ANNLO}
\ANNLO(p,p,E) = -\frac{\pi \rho_d^2}{m_N} \frac{ (\gamma-ip)^2}{\gamma+ip}~.
\end{equation}
The $\mu$-dependence of Eq.~\eqref{eq:NNLOhalf} is removed in the 
on-shell amplitude of Eq.~\eqref{eq:ANNLO}.

\subsection{Deuteron electric form factor}
In the vicinity of the bound-state pole $p =i\gamma$, the on-shell amplitude is given by
\begin{eqnarray}
\label{eq:Ad}
\mathcal{A}_{d}(E) &=&  - \frac{\cZd}{E+\gamma^2/m_N}~, 
\end{eqnarray}
where $\cZd$ is the wave function renormalization factor 
related to $\rho_d$ and $\gamma$ as
\begin{eqnarray}
  \label{eq:Zdfull}
  \cZd &=&  \frac{8\pi\gamma}{m_N^2(1-\rho_d \gamma)}~.
\end{eqnarray}
Eqs.~\eqref{eq:Ad} and \eqref{eq:Zdfull} are needed in the calculation
of the deuteron electric form factor that has been calculated up to 
NNLO in $\slashed{\pi}$EFT as~\cite{Chen:1999tn}
\begin{equation}
\label{eq:FE}
F_{\text{E}}(q^2)=\frac{1}{1-\rho_d \gamma}\left[\frac{4\gamma}{q}
\arctan\frac{q}{4\gamma}-\rho_d \gamma\right]\,.
\end{equation}
The form factor of Eq.~\eqref{eq:FE} is needed in evaluating the  
third Zemach moment contribution to the Lamb Shift in muonic deuterium
given by Eq.~\eqref{eq:Zem}.

\section{Diagrammatic calculation of transition matrix elements}
\label{sec:diagr-calc-struct}
In this section, we use $\slashed{\pi}$EFT Feynman diagrams to
evaluate the longitudinal transition matrix elements up to its NNLO
contribution, which is needed for the evaluation of the longitudinal
structure function and the polarizability effect.

\subsection{Transition matrix element at LO}
\begin{figure}[t]
\begin{center}
\includegraphics[width=0.9\textwidth,clip=true]{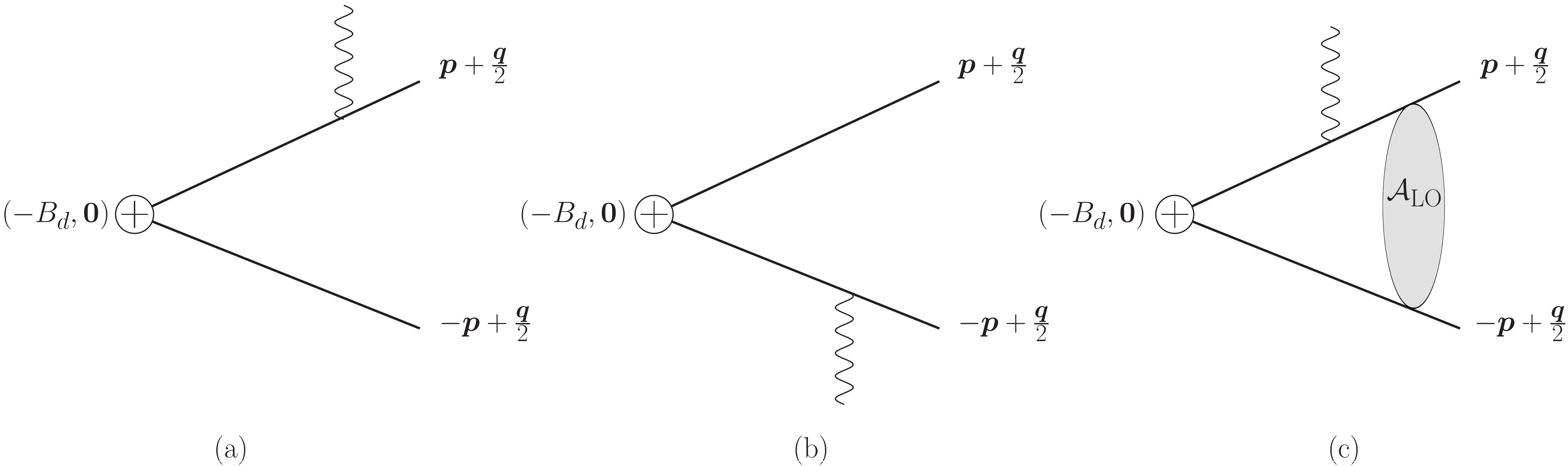} 
\end{center}
\caption{Contribution to the leading order transition matrix element
  needed for the inelastic longitudinal structure function. The wavy
  line denotes the coupling of the electromagnetic current to the
  nucleon.}
\label{fig:MLO}
\end{figure}
At LO, the transition matrix element from the $A_0$-photon excitation
of the deuteron is depicted by the Feynman diagrams in
Fig.~\ref{fig:MLO}. The first two diagrams, (a) and (b), are
contributions without final state interactions, while the third diagram
contains final state interactions as indicated by a shaded grey oval 
symbolizing an insertion of the LO scattering amplitude.
Figure~\ref{fig:MLO} also displays the kinematics we use in the 
calculation of this matrix element. Note that the incoming
deuteron is at rest, so the total energy of the initial two-nucleon 
state is therefore $-B_d$. The final two-nucleon 
state has energy $E=p^2/m_N$.

We evaluate the plane-wave contribution by calculating the amplitude
without projecting on specific spin-orbit coupled final states and are
thereby summing up all electric multipoles contributing to the
transition induced by the $A_0$ photon.  The plane-wave final states
are represented by two nucleon spinors $N^T$ and $N$ for each outgoing
nucleon leg. $N$ holds a tensor product of the two-component spinor
for spin and the two-component spinor for isospin. The initial
deuteron $^3{\rm S}_1$ state is projected out by the operator
$P_i$. The antisymmetrization is treated by including both diagrams
where the photon either couples to the top or the bottom leg as shown
in diagrams (a) and (b) of Fig.~\ref{fig:MLO}.

For diagram (a) in Fig.~\ref{fig:MLO}, we obtain
\begin{align}
\label{eq:MLOa}
  \nonumber
(i \mathcal{M}_{\rm LO}^{a})^{i}_{\alpha\beta}  &= i 2 \sqrt{\cZd}  i S({\textstyle -B_d - \frac{(-\bs{p}+\bs{q}/2)^2}{2m_N}, \bs{p} - \frac{\bs{q}}{2}}) 
\, N_\alpha^T(\bs{p}+\frac{\bs{q}}{2})\mathcal{Q}P_i^\dagger
                   N_\beta(-\bs{p}+\frac{\bs{q}}{2})~\\
  & = i \sqrt{\cZd} i {\tilde{M}_{\rm LO}^{a}} N_\alpha^T({\bf
  p}+\frac{\bs{q}}{2})\mathcal{Q}P_i^\dagger
                   N_\beta(-\bs{p}+\frac{\bs{q}}{2}).
\end{align}
where we defined the amplitude
${\tilde{M}_{\rm LO}^{a}} = -2m_N /\left[ \gamma^2 +
  (\bs{p}-\bs{q}/2)^2\right]$.  The indices $\alpha$ and $\beta$ on
the nucleon spinors denote spin and isospin basis indices, and $i$ is
the deuteron spin index. Note that a factor of 2 is included in
Eq.~\eqref{eq:MLOa} to account for both possible contractions leading 
to that expression. Similarly, diagram (b) in Fig.~\ref{fig:MLO}~yields
\begin{align}
(i \mathcal{M}_{\rm LO}^{b})^i_{\alpha\beta}
  = i \sqrt{\cZd}i {\tilde{M}_{\rm LO}^{b}} N_\alpha^T({\bf
  p}+\frac{\bs{q}}{2})P_i^\dagger\mathcal{Q}
  N_\beta(-\bs{p}+\frac{\bs{q}}{2})\,,
\end{align}
where the amplitude
${\tilde{M}_{\rm LO}^{b}} = -2m_N /\left[ \gamma^2 +
  (\bs{p}+\bs{q}/2)^2\right]$.

Furthermore, we evaluate the contributions with $^3{\rm S}_1$
final-state interactions at LO, using the diagram (c) shown in
Fig. \ref{fig:MLO}. The excitation from $^3{\rm S}_1$ bound state to
the $^1{\rm S}_0$ scattering state is forbidden by the $A_0$
photon. The iterative sum of final-state interactions is represented
by the off-shell scattering amplitude $P_i^\dagger i \ALO P_i$.

Using the Feynman rules we obtain for this diagram the transition matrix element
\begin{multline}
  \label{eq:lo-with-fsi}
 ( i \mathcal{M}_{\rm LO}^{c})^i_{\alpha \beta}= i 8 \sqrt{\cZd} \:\hbox{Tr}\left[P_i^\dagger  \mathcal{Q} P_j\right]   
  i \ALO(E) 
  \,N_\alpha^T(\bs{p}+\frac{\bs{q}}{2}) P^\dagger_j N_\beta(-\bs{p}+\frac{\bs{q}}{2}) \\
\times\int \frac{\hbox{d}^4 l}{(2\pi)^4} iS(-B_d+l_0,\bs{l}-\frac{\bs{q}}{2})
i S(-B_d+l_0+\omega, \bs{l}+\frac{\bs{q}}{2}) i S(-l_0,\bs{l}-\frac{\bs{q}}{2})~.
\end{multline}
Note that a factor of 8 is included for the possible number of
contractions in the matrix element evaluation.
The loop integral arising in the calculation
of diagram (c) in Fig.~\ref{fig:MLO} we label as $\mathcal{J}_0$, and
after evaluating the trace
$\hbox{Tr}[P_i^\dagger \mathcal{Q} P_j] = \delta_{ij}/4$ 
in Eq.~\eqref{eq:lo-with-fsi}, we find
\begin{align}
  \label{eq:Mc-short}
  (i \mathcal{M}_{\rm LO}^{c})^i_{\alpha \beta}
  & = i\sqrt{\cZd} i \tilde{M}_{\rm LO}^{c} N_\alpha^T(-\bs{p}+\frac{\bs{q}}{2}) P^\dagger_i N_\beta(\bs{p}+\frac{\bs{q}}{2})~,
\end{align}
where $ {\tilde{M}_{\rm LO}^{c}} = 2 \cJ_0 \ALO(E) $. The definition
and the derivation of the $\cJ_0$ integral result are given in
App.~\ref{sec:J0}.

Below, at NLO and NNLO, we will use the short-hand notation 
introduced above and work
with the amplitudes $\tilde{M}$ with $\sqrt{Z_d}$ and the spinor
pieces factored out. Amplitudes with superscripts $a$ or $b$ 
will continue to correspond to diagrams without final-state 
interactions, while those with superscript $c$ correspond to 
diagrams at a given order that contain final-state interactions.

\subsection{Transition matrix element at NLO}
\label{sec:inel-struct-funct}
\begin{figure}[t]
\begin{center}
\includegraphics[width=0.8\textwidth,clip=true]{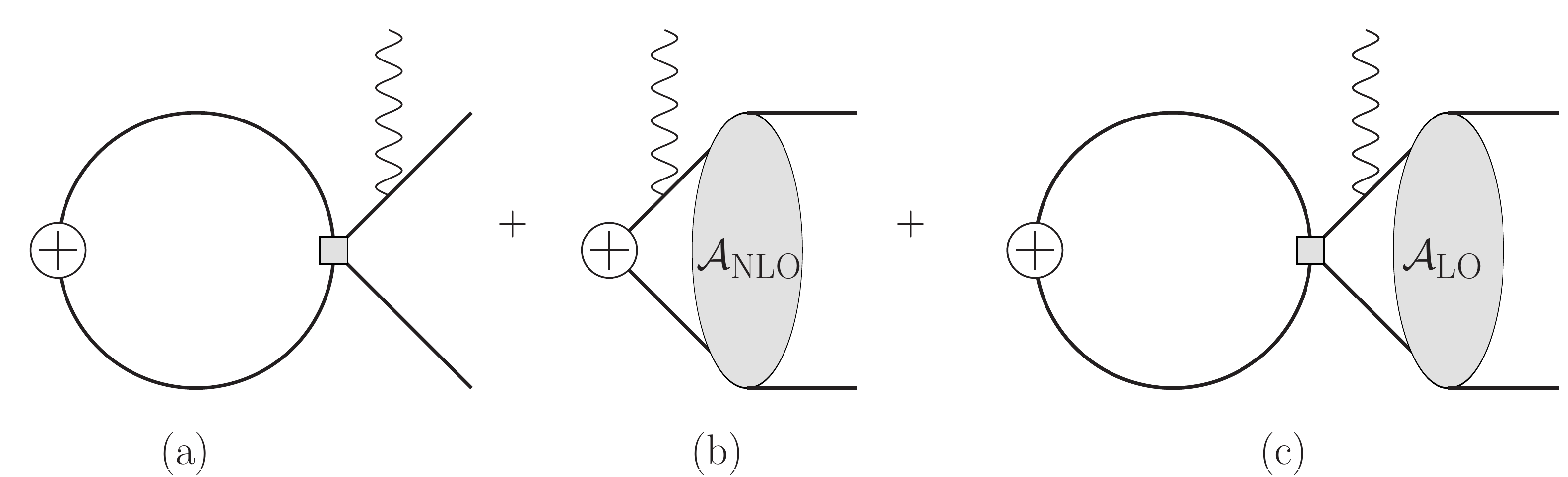} 
\end{center}
\caption{\label{fig:MNLO}NLO diagrammatic contribution to the matrix
  element $\mathcal{M}$. Final state S-wave interactions are indicated by
  the grey blob. We have omitted one diagram without final state
  interactions where the photon couples to the lower outgoing leg.}
\end{figure}

Diagrams that contribute to the transition matrix element at NLO are shown in Fig.~\ref{fig:MNLO}.
The diagram with one NLO contact that has no final state interactions 
is given by diagram (a) of Fig.~\ref{fig:MNLO}. After
antisymmetrization, it leads to the expressions 
\begin{align}
\label{eq:MaNLO}
  \nonumber
  i \tilde{M}_{\rm NLO}^{a}  =
&  2 \left [(-i \frac{C_{2,-2}}{2})
 \left(i\cI_0(-B_d) \left(\textstyle{\bs{p}-\frac{\bs{q}}{2}}\right)^2
 + i \cI_2(-B_d)\right)
 +(-iC_{0,0})i \cI_0(-B_d)\right]\\
&\qquad\qquad \times i S(\textstyle{-B_d - \frac{(-\bs{p}+\bs{q}/2)^2}{2m_N}, 
	\bs{p} - \frac{\bs{q}}{2}}) ~\nn\\
 =& i \frac{\rho_d}{2} m_N (\mu - \gamma)^{-1}~
\end{align}
and
\begin{equation}
	\label{eq:MbNLO}
	i\tilde{M}_{\rm NLO}^{b} = i\frac{\rho_d}{2} m_N (\mu - \gamma)^{-1}~,
\end{equation}
where we used that
$-\cI_2(-B_d)=\gamma^2 \cI_0(-B_d)$ and $C_{0,0} = \gamma^2 C_{2,-2} $.

Diagram (b) of Fig.~\ref{fig:MNLO}~gives the contribution with no 
explicit contact term insertion, but with one NLO final-state
scattering amplitude.\footnote{The diagram label (b) in Fig.~\ref{fig:MNLO}
is not connected with the superscript $c$ indicating final-state
interactions.} It leads to
\begin{align}
  \label{eq:Mc0-2}
  i \tilde{M}_{\rm NLO}^{c,0} =& 2 \frac{\rho_d}{2} i\ALO(E)  \int
                            \frac{\hbox{d}^3 k}{(2\pi)^3}
                            i S(-B_d-\frac{k^2}{2m_N}, \bs{k})
                                  i S(-B_d+\omega-\frac{k^2}{2m_N}, \bs{k}+\bs{q})\nn\\
                           &\times
                            \Biggl[(\gamma-i p)+\frac{1}{2(\gamma-\mu)}(v^2-p^2)\Biggr]\,,
\end{align}
where $v = |{\bf k}+\frac{\bf q}{2}|$ and the half off-shell NLO 
amplitude $\ANLO(v,p,E)$ of Eq.~\eqref{eq:ANLO1} depending on the loop 
momentum $k$ was inserted. Using the definitions from the appendix for the loops that 
couple to the photon, we can write
\begin{align}
  \label{eq:Mc0-4}
  i\tilde{M}_{\rm NLO}^{c,0} &=  2
  i\ALO(E)  \frac{\rho_d}{2}\biggl[ (\gamma - i p) \cJ_0 
+\frac{1}{2(\gamma-\mu)}\left(\tilde{\cJ}_2 - p^2 \cJ_0\right)\biggr]~\nn\\
&=\frac{\rho_d}{2}i\ALO(E)\biggl[ 2 (\gamma - i p) \cJ_0 
+\frac{m_N^2}{4\pi}\biggr]~,
\end{align}
where $\tilde{\cJ}_2$ is defined in Eq.~\eqref{eq:J2bar} and we used
the result of Eq.~\eqref{eq:J2bar-sum-1} in App.~\ref{sec:J2} to
replace $\left(\tilde{\cJ}_2 - p^2 \cJ_0\right)$.  Diagram (c) of
Fig.~\ref{fig:MNLO}~with one NLO operator and one LO scattering
amplitude yields an amplitude we label $i \tilde{M}_{\rm
  NLO}^{c,1}$. Combining the sum of loop integrals and using the
identities $C_{0,0} = \gamma^2 C_{2,-2}$ and
$\cI_2(-B_d) = -\gamma^2 \cI_0 (-B_d)$ allows for this amplitude to be
written
\begin{align}
  \label{eq:Mbar-c1NLO}
 \tilde{M}_{\rm NLO}^{c,1}&= C_{2,-2} \cI_0(-B_d) (\cJ_2+\gamma^2\cJ_0) 
 \ALO(E) \nn\\
 &= - \frac{\rho_d}{2}\frac{m_N}{\mu-\gamma}\cI_0(E)   \ALO(E)~,
\end{align}
where the expressions for loop integrals from
Eq.~\eqref{eq:J2-gJ0-pds} are used. The summation of Eqs.~\eqref{eq:Mc0-4}
and \eqref{eq:Mbar-c1NLO} yields
\begin{align}
\label{eq:Mbar-cNLO}
\tilde{M}_{\rm NLO}^c =& \tilde{M}_{\rm NLO}^{c,0} + 
\tilde{M}_{\rm NLO}^{c,1}~,\nn\\
=& \frac{\rho_d}{2} \ALO(E) \left[ 2 (\gamma-ip)\cJ_0  + \frac{m_N^2}{4\pi} \left(1+\frac{\mu+ip}{\mu-\gamma}\right)\right]~.
\end{align}

\subsection{Transition matrix element at NNLO}
\label{sec:inel-struct-funct-1}
\begin{figure}[t]
\begin{center}
\includegraphics[width=0.95\textwidth,clip=true]{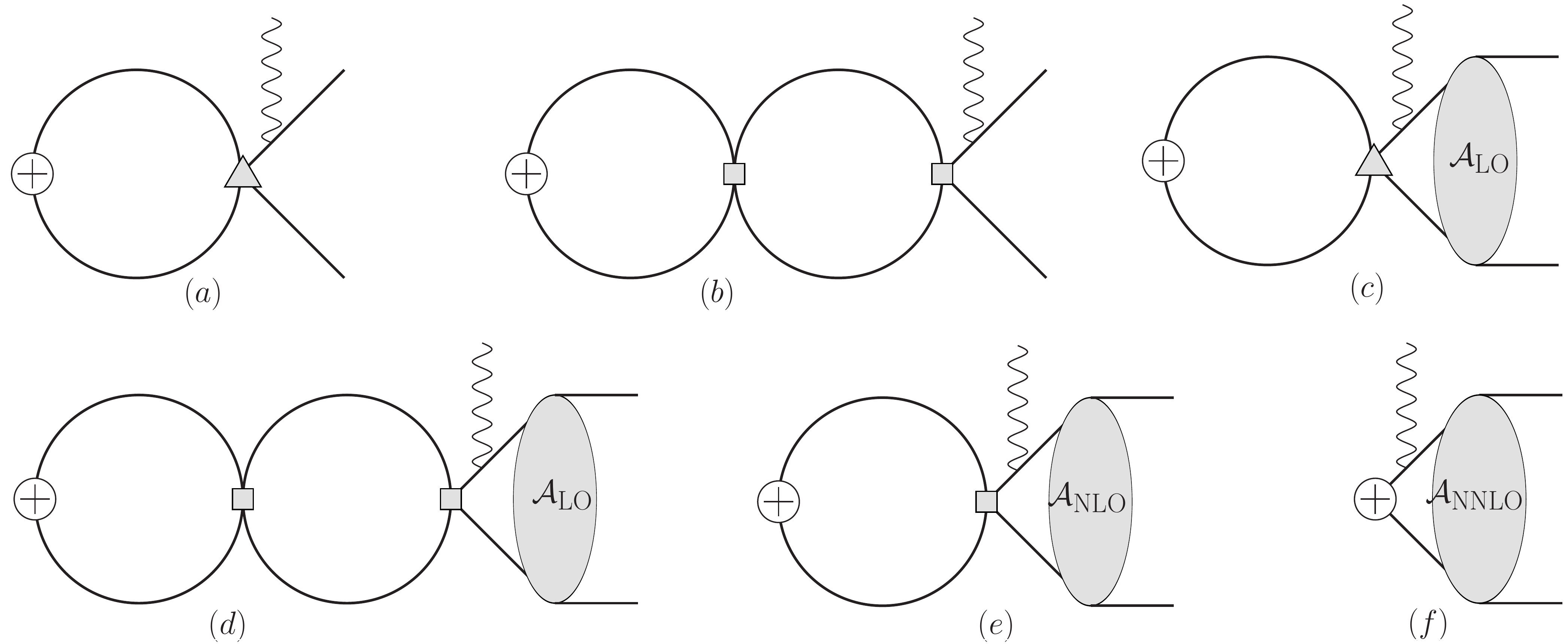} 
\end{center}
\caption{\label{fig:M-n2lo-0}NNLO diagrammatic contribution to the
  matrix element $\mathcal{M}$ that includes one insertion of NNLO
  operator or two insertions of NLO operator. The grey square and
  triangle denote the collection of NLO and NNLO operators,
  respectively. These diagrams make zero contributions after
  renormalization.}
\end{figure}
In Fig.~\ref{fig:M-n2lo-0}, we show the diagrams contributing at
NNLO. Diagrams (a) and (b) in Fig.~\ref{fig:M-n2lo-0} generate the
NNLO contributions to $\cM^a$ and $\cM^b$. The sum of these diagrams
give no contribution because
$\tilde{M}_{\rm NNLO}^{a}=\tilde{M}_{\rm NNLO}^{b}=0$. Diagrams (c)
and (d) in Fig.~\ref{fig:M-n2lo-0} evaluate to zero, too.  Diagrams
(e) and (f) in Fig.~\ref{fig:M-n2lo-0} represent the non-zero
contributions to the transition matrix element at NNLO. Diagram (e)
gives the contributions with one insertion of one NLO operator and one
NLO amplitude. Its contribution to $\cM^c$ yields
\begin{align}
   \label{eq:n2lo_fsi_1_1}
    i\tilde{M}_{\rm NNLO}^{c,0}
    &= C_{2,-2} \cI_0(-B_d) (\cJ_2 + \gamma^2 \cJ_0) \frac{\rho_d}{2} i\ALO(E) (\gamma-ip)~,
    \nn\\
    & = i\ALO(E) \left(\frac{\rho_d}{2}\right)^2 \frac{m_N^2}{4\pi} (\gamma-ip)\frac{\mu+ip}{\mu-\gamma} ~,
\end{align}
where the expression of loop integral in Eq.~\eqref{eq:J2-gJ0-pds} is
used.  Inserting the half-off-shell NNLO amplitude in diagram (f), we
obtain
\begin{align}
  \label{eq:Mc1N2LO}
  i \tilde{M}_{\rm NNLO}^{c,1} &=
  2 i \ALO(E)\left(\frac{\rho_d}{2}\right)^2\left[
    (\gamma - i p)^2 \cJ_0
    +\frac{\gamma - ip}{2(\gamma-\mu)}
    (\tilde{\cJ}_2-p^2 \cJ_0)(1+\frac{\gamma+i p}{\gamma - \mu})
    \right]~\nn\\
  &=2 i \ALO(E)\left(\frac{\rho_d}{2}\right)^2
  \biggl[(\gamma- ip )^2 \cJ_0 +\frac{m_N^2}{8\pi}(\gamma - ip)
(1+\frac{\gamma+ip}{\gamma -\mu})
  \biggr]~.
\end{align}
The regulator-dependence in the NNLO contribution is removed when
Eqs.~\eqref{eq:n2lo_fsi_1_1} and \eqref{eq:Mc1N2LO} are summed
together as
\begin{equation}
  \label{eq:McN2LO}
   \tilde{M}_{\rm NNLO}^{c} =2  \ALO(E)\left(\frac{\rho_d}{2}\right)^2(\gamma- ip )
  \biggl[(\gamma- ip ) \cJ_0 +\frac{m_N^2}{4\pi}\biggr]~.
\end{equation}

We note that the S-D wave mixing operator enters at NNLO in the
$\slashed{\pi}$EFT Lagrangian, and in principle gives a contribution to the
transition matrix element at the same order. However, due to the
orthogonality of S-wave and D-wave component, contributions from the
D-wave projection to the squared matrix element does not interfere
with the S-wave projection. Therefore, the S-D mixing contributions
enter at N4LO in the squared matrix element, and we
therefore do not consider this higher-order effect.

\subsection{Matrix element squared}
The calculation of the inelastic longitudinal structure function
in Eq.~\eqref{eq:SL-def} requires the squared matrix elements and a 
sum over the outgoing nucleon-nucleon spin and isospin states. This 
sum leads to traces of products of projection and charge operators.  
We carry this out without projecting the outgoing
state on a specific spin or isospin coupling. We write
\begin{align}
\label{eq:Msq}
\overline{|\cM|^2}&=\frac{1}{2}\sum_{\alpha \beta} |
                \mathcal{M}_{\alpha\beta}^a+\mathcal{M}_{\alpha
                \beta}^b+ \mathcal{M}_{\alpha\beta}^c|^2~,
\end{align}
where the factor of $1/2$ is introduced to account for the
identicality of particles in the intermediate state. The summation
over outgoing spins converts the included spin-projected matrix
elements into traces over the corresponding projectors. After
evaluating the traces arising in Eq.~\eqref{eq:Msq}, we obtain
\begin{align}
  \label{eq:Msq2}
  \overline{|\cM|^2} = \frac{\cZd}{4} \left( \left|\frac{{\tilde{M}^a}-{\tilde{M}^b}}{2}\right|^2 + \left|\frac{{\tilde{M}^a}+{\tilde{M}^b}}{2}+{\tilde{M}^c}\right|^2\right)~,
\end{align}
where the amplitudes on the right hand side can be expanded 
order-by-order. The first term in Eq.~\eqref{eq:Msq2} receives 
no NLO or NNLO
corrections since $\tilde{M}_{\rm NLO}^{a}-\tilde{M}_{\rm NLO}^{b}= 0$,
as can be seen from Eqs.\eqref{eq:MaNLO} and \eqref{eq:MbNLO},
and $\tilde{M}_{\rm NNLO}^a = \tilde{M}_{\rm NNLO}^b =0$.

The second term in Eq.~\eqref{eq:Msq2} receives LO, NLO, and NNLO 
contributions, but the NNLO contribution 
arises only from $\tilde{M}_{\rm NNLO}^c$,
given in Eq.~\eqref{eq:McN2LO}. We can therefore
write the squared amplitude up through NNLO as
\begin{align}
  \label{eq:NNLOsum}
  \nonumber
  \overline{|\cM|^2} =&\frac{ \cZd }{4}\Biggl\{
  \left| \frac{\tilde{M}_{\rm LO}^a-\tilde{M}_{\rm LO}^b}{2}\right|^2
     +\Biggl| \frac{\tilde{M}_{\rm LO}^a+\tilde{M}_{\rm LO}^b}{2} 
     \nonumber\\
     &\quad +2\ALO(E)\left[\cJ_0\left(
     \sum_{n=0}^{2}\left(\frac{\rho_{d}}{2}\right)^{n}
     \left(\gamma-ip\right)^{n}   \right)
     +\frac{\rho_d}{2}\frac{m_N^2}{4\pi}\left(1+
     \frac{\rho_d}{2} (\gamma - ip)\right)\right]
    \Biggr|^2\Biggr\}~.
\end{align}

\section{Results}
\label{sec:Results}
Analytical results for the inelastic longitudinal structure function
can be calculated by inserting Eq.~\eqref{eq:NNLOsum} into the
integrand in Eq.~\eqref{eq:SL-def}. Once the structure function is
obtained at each order we consider in this work, we compute
$\delta_{\text{pol},L}^{A}$ order-by-order. We extract the electric
dipole contribution for comparison with other works that carry out an
explicit multipole
decomposition~\cite{Leidemann1995,Hernandez:2019zcm}. Additionally, we
calculate the third inelastic Zemach moment term
$\delta_{{\rm Zem}}^{A}$.

\subsection{Results for the longitudinal structure function}
To demonstrate the order-by-order convergence of the
$\slashed{\pi}$EFT calculation of the longitudinal structure function
$S_L$, we extract the $\rho_d \gamma$ power dependence at each order.
Besides the explicit $\rho_d \gamma$ dependence in
Eq.~\eqref{eq:NNLOsum}, we expand the the deuteron renormalization
constant $\cZd$ in powers of $\rho_d \gamma$ as
\begin{equation}
\label{eq:Zexpand}
 \cZd = \frac{8\pi \gamma}{m_N^2} 
 \left( 1+ \rho_d \gamma + \rho_d^2 \gamma^2+\cdots\right),
\end{equation}
to include the order-by-order correction from the wave function
renormalization.

At leading order we find the inelastic structure function result
\begin{align}
\label{eq:SLLO}
S_{L}^{\text{LO}}\left(q,\omega\right)
=&
\frac{\gamma p}{\pi m_N}
\bigg[\frac{2m_N^2}{m_N^2\omega^2-q^2 p^2}+
\frac{32\pi^2}{m_N^2(\gamma^2+p^2)}\nn\\
&\qquad \times\Big(\big(\text{Re}\big[\cJ_{0}\big]\big)^2-
\frac{2\gamma}{p}\text{Re}\big[\cJ_{0}\big]
\text{Im}\big[\cJ_{0}\big]-\big(\text{Im}\big[\cJ_{0}\big]\big)^2\Big)
\bigg]\,,
\end{align}
where $p=\sqrt{m_N\omega-\gamma^2-q^2/4}$.
At NLO, the inelastic longitudinal structure function we calculate as
\begin{align}
\label{eq:SLNLO}
S_{L}^{\text{NLO}}(q,\omega)=&\rho_{d}\gamma S_{L}^{\text{LO}}(q,\omega)+
\frac{8 \rho_d \gamma}{m_N(\gamma^2+p^2)}
\bigg\{ 
p\text{Re}\left[\cJ_{0}\right]-\gamma \text{Im}\left[\cJ_{0}\right]
\nn\\
&+\frac{4\pi}{m_N^2} \Big[
\gamma p \left(\left(\text{Re}\left[\cJ_{0}\right]\right)^2
-\left(\text{Im}\left[\cJ_{0}\right]\right)^2\right)
+(p^2-\gamma^2) \text{Re}\left[\cJ_{0}\right] \text{Im}\left[\cJ_{0}\right] 
\Big]\bigg\}\,.
\end{align}
Lastly, the N2LO part of the inelastic structure function is 
\begin{align}
\label{eq:SLN2LO}
S_{L}^{\text{NNLO}}=&\rho_{d}\gamma S_{L}^{\text{NLO}}(q,\omega)
+\frac{m_N\gamma\rho_{d}^{2}}{2\pi(\gamma^2+p^2)}
\bigg\{ p
+ \frac{8\pi}{m_N^2}\Big[
2\gamma p \text{Re}\left[\cJ_{0}\right]
+ (p^2-\gamma^2)\text{Im}\left[\cJ_{0}\right]\Big]
\nn\\
&+\frac{16\pi^2}{m_N^4} \Big[ 
(p^3-3p\gamma^2) \left( \text{Im}\left[\cJ_{0}\right]^2-\text{Re}\left[\cJ_{0}\right]^2\right)
+(6p^2\gamma-2\gamma^3) \text{Re}\left[\cJ_{0}\right] \text{Im}\left[\cJ_{0}\right]
\Big]
\bigg\}\,.
\end{align}

\begin{figure}
	\begin{center}
		\includegraphics[width=0.47\textwidth,clip=true]{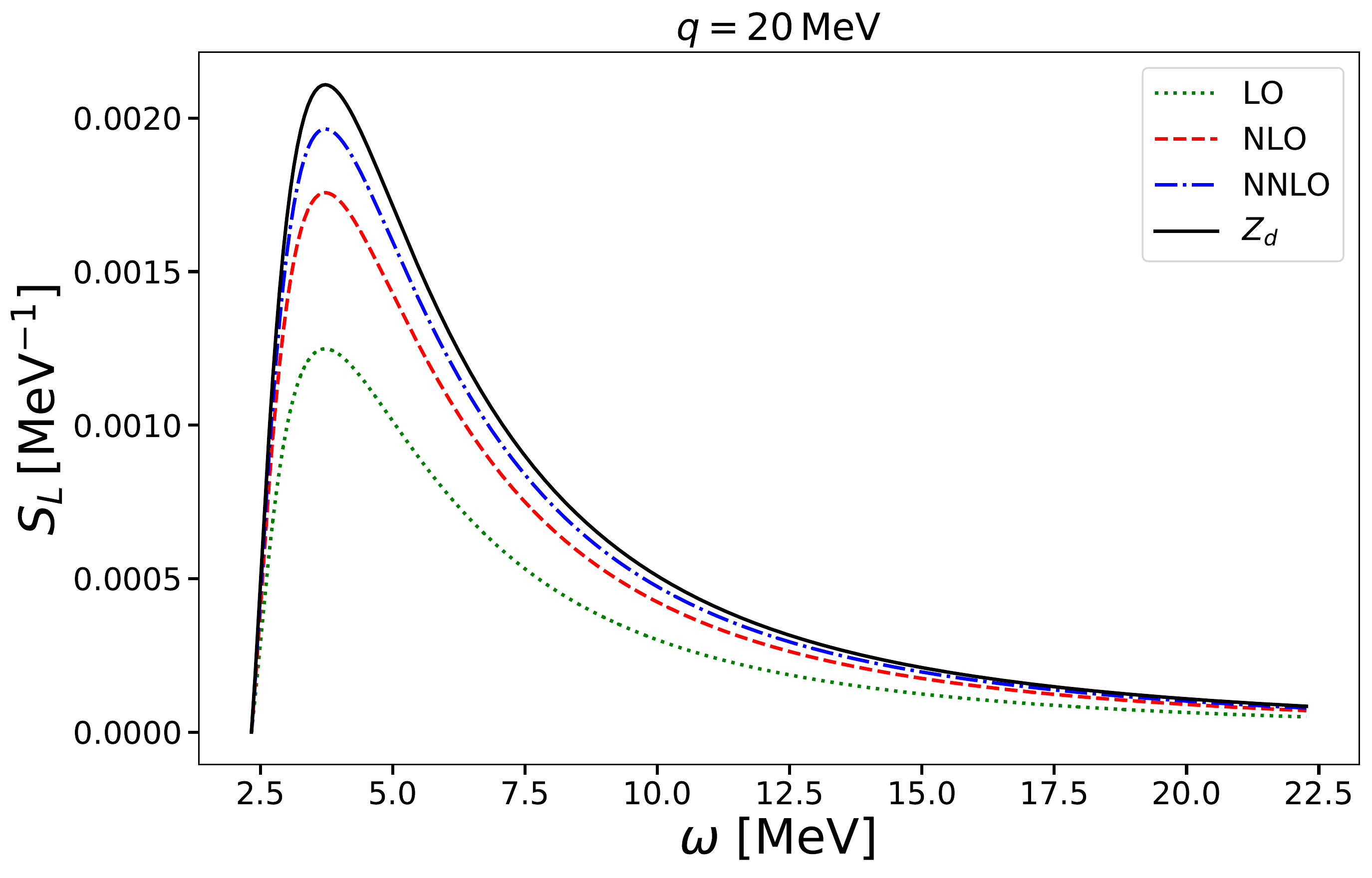}
		\includegraphics[width=0.47\textwidth,clip=true]{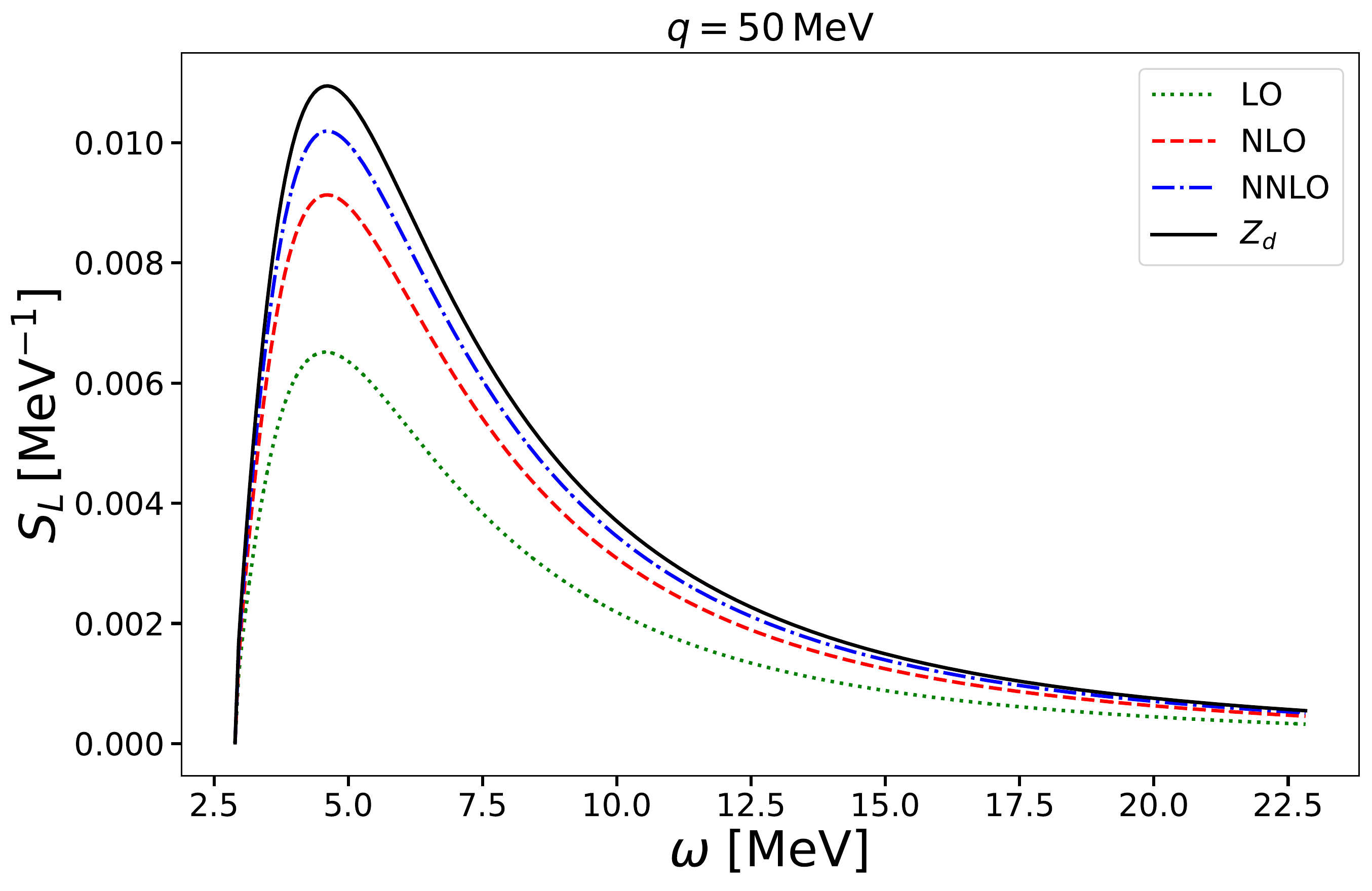}
	\end{center}
	\caption{\label{fig:SL} Longitudinal structure function $S_L$
          as a function of $\omega$ at fixed momentum exchange
          $q = 20$~MeV (left panel) and $q= 50$~MeV (right panel). The
          dotted, dashed, dot-dashed lines give the result of the LO,
          NLO and NNLO structure function, respectively. The solid
          line gives NNLO $\mathcal{Z}_d$-improved result.}
\end{figure}

The EFT convergence of the structure function is
shown in Fig.~\ref{fig:SL}, where $S_L(\omega,q)$ is plotted as a
function of $\omega$ by fixing $q$ at 20 MeV and 50 MeV
in the two different plots in the figure. 
Calculations of $S_L(\omega,q)$ at LO, NLO, and NNLO are
compared in the plots, which show an order-by-order convergence in
$\slashed{\pi}$EFT.

\subsection{Benchmark with dipole approximation results}
As mentioned previously, we do not perform our calculation using a
multipole decomposition as in Refs.~\cite{Leidemann1995,Hernandez:2019zcm}.  
Instead, the integration to determine $\delta_{{\rm pol},L}^{A}$ in
Eq.~\eqref{eq:polL} was taken using the complete inelastic
longitudinal structure function of the nucleus, given 
order-by-order in Eqs.~\eqref{eq:SLLO}, \eqref{eq:SLNLO}, 
and $\eqref{eq:SLN2LO}$, that implicitly
contains all inelastic multipole contributions. However,
we extract the inelastic dipole excitation for comparison
with previous literature. This is motivated by the fact that the
dipole excitation gives the largest contribution to TPE.

The electric-dipole excitation of the nucleus arises when the outgoing
unbound nucleons are in a spin-singlet isospin-triplet state. The
antisymmetry of the wave function is preserved by the fact that the
outgoing $NN$ state has odd orbital angular momentum. The first term
in the low-momentum approximation of the squared matrix element in
Eq.~\eqref{eq:Msq2} corresponds to a $P$ wave between the
outgoing nucleons and is
\begin{align}
  \label{eq:me_dipole}
  \overline{|\cM|^2} \approx  \frac{\cZd}{4}  \left|\frac{\tilde{M}^a_{\rm LO}-\tilde{M}^b_{\rm LO}}{2}\right|^2 
  \approx m_N^2 \cZd \frac{(\bs{p}\cdot \bs{q})^2}{(\gamma^2+p^2)^4}~.
\end{align}
Inserting the matrix element $\overline{|\cM|^2}$ into a low-$q$
truncated version of Eq.~\eqref{eq:SL-def} by dropping the recoil energy $q^2/4m_N$ gives the dipole part of
the inelastic longitudinal structure function as
\begin{align}
\label{eq:SLDLO}
 S_{D}(|\bm{q}|,\omega) =& \int \frac{\hbox{d}^3 p}{(2\pi)^3}
  \delta(\omega-\frac{\gamma^2}{m_N}- \frac{p^2}{m_N} )\, 
  \overline{|\cM|^2}
  \nn \\
 =&\frac{\cZd \sqrt{m_N}}{12\pi^2}
\frac{\left(\omega-\gamma^2/m_N\right)^{3/2}}
{\omega^4}q^2\,.
\end{align}
Plugging the dipole structure function $S_{D}(|\bm{q}|,\omega)$ of
Eq.~\eqref{eq:SLDLO} and the non-relativistic kernel $K_{L}^{NR}$ from
Eq.~\eqref{eq:KLNR} into the TPE energy-shift equation given by 
Eq.~\eqref{eq:polL} yields
\begin{equation}
\label{eq:dipole}
\delta_{{\rm pol},D}^A = -\frac{4\pi}{3}\alpha^2\phi^2(0) \frac{\cZd m_N^{5/2} \sqrt{2m_r}}{35\pi^2 \gamma^4}~,
\end{equation}
which matches the expression of the same contribution in
Ref.~\cite{Friar2013}. In the above expression, we replaced $m_\mu$ with the $\mu$-d reduced mass $m_r$ to adjust the truncation of recoil correction in
the low-q approximation. The numerical evaluation of the dipole
term given by Eq.~\eqref{eq:dipole}
yields $\delta_{{\rm pol},D}^A = -1.925$~meV.

\subsection{Numerical results for TPE contribution to Lamb shift}
The TPE contribution to the Lamb shift in muonic deuterium can be
calculated order-by-order in $\slashed{\pi}$EFT to the desired
precision. We do so up to NNLO in an expansion in the parameter
$\rho_{d}\gamma$. To evaluate the elastic TPE, Zemach term
$\delta^A_{\text{Zem}}$, we insert the electric form factor
$F_{\rm E}$ from Eq.~\eqref{eq:FE} into the integral
equation~\eqref{eq:Zem}. $\delta^A_{\rm Zem}$ is evaluated to be
$-0.362\,\text{meV}$, which is consistent with the calculation in
Ref.~\cite{Hernandez:2019zcm}.

\begin{table}[t] 
	\begin{tabular*}{0.6\textwidth}{@{\extracolsep{\fill}} l c c}
		$\delta_{{\rm pol},L}^A$ & NR limit & 
		R limit \\
          \hline
		$\slashed{\pi}$EFT LO                  & -0.962 & -0.943 \\
		$\slashed{\pi}$EFT NLO                 & -1.346 & -1.320 \\
		$\slashed{\pi}$EFT NNLO                & -1.499 & -1.470 \\
		$\mathcal{Z}_d$ improved               & -1.605 & -1.574 \\
		dipole term                            & -1.925 &  --    \\		
		ZRA ($\eta$-less)                      & -1.590 & -1.553 \\
		ZRA ($\eta$-expansion)                 & -1.590 & -1.564 \\
		$\chi$EFT ($\eta$-less)                & -1.588 & -1.562 \\
		$\chi$EFT ($\eta$-expansion)           & -1.590 & -1.560 \\
        \hline
	\end{tabular*}
	\caption{The longitudinal polarizability $\delta_{{\rm
              pol},L}^A$ is calculated in both non-relativistic and
          relativistic limit. Results in $\slashed{\pi}$EFT are
          calculated in this work at LO, NLO, NNLO and with the NNLO
          $\cZd$-improvement approach. Results from other works in
          dipole approximation, ZRA and $\chi$EFT are extracted from
          information in
          Refs.~\cite{Friar2013,Hernandez:2019zcm,Ji:2018ozm}.}
	\label{tab:Shifts}
\end{table}

The longitudinal polarizability $\delta^A_{\text{pol},L}$ is firstly
evaluated in the non-relativistic approximation by using the
non-relativistic kernel $K_{L}^{NR}$ from Eq.~\eqref{eq:KLNR} in the
TPE sum rule.  At leading order in $\slashed{\pi}$EFT expansion, we
obtain
\begin{equation}
\label{eq:deltaLO}
[\delta^A_{\text{pol},L}]_{\text{LO}}=-0.962\,\text{meV}\,.
\end{equation}
The NLO correction to $\delta^A_{\text{pol},L}$ consists of two parts, whose sum gives
\begin{equation}
\label{eq:deltaNLO}
\Delta [\delta^A_{\text{pol},L}]_{\text{NLO}}=(-0.393 + 0.009)\, \text{meV}~.
\end{equation}
The dominant contribution arising from $\rho_d\gamma$ expansion in the
$\cZd$ factor corresponding to the first term in
Eq.~\eqref{eq:deltaNLO} is given by
$\rho_d \gamma [\delta^A_{\text{pol},L}]_{\text{LO}}$, and the rest is
due to the NLO diagrams in Fig.~\ref{fig:MNLO}.  The dominant
contribution to the NNLO correction on $\delta^A_{\text{pol},L}$
arises from the $\rho_d \gamma$ expansion in $\cZd$, and is given by
$\rho_d \gamma \Delta [\delta^A_{\text{pol},L}]_{\text{NLO}}$. The
NNLO diagrams in Fig.~\ref{fig:M-n2lo-0} lead to a contribution that
is signicantly smaller. The sum of the two NNLO contributions yields
\begin{equation}
\label{eq:deltaNNLO}
\Delta [\delta^A_{\text{pol},L}]_{\text{NNLO}}=(-0.157 + 0.004)\, \text{meV}~.
\end{equation}
By summing up the contributions at each order, we obtain
$\delta^A_{\text{pol},L}$ to be $-0.962$ meV, $-1.346$ meV and
$-1.499$ meV at LO, NLO and NNLO respectively.

The $(\rho_d \gamma)$-expansion of $\cZd$ clearly dominates the
effective range corrections to $\delta^A_{\text{pol},L}$ and effects
of diagrams that include final state interactions beyond NNLO are
suppressed. The accuracy of the calculation can therefore be improved
above NNLO by simply using the resummed wave function renormalization
$\cZd$ in the evaluation\footnote{Note that this is not Z-matching as
  introduced in Ref.~\cite{Phillips:1999hh} since we do not change the
  low-energy coefficients used in the evaluation of the transition
  matrix element.}. We first separate out the diagrammatic
contributions at LO, NLO and NNLO respectively by setting
$\cZd$ equaling to its LO value $8\pi \gamma/m_N^2$. The sum of the results are
then multiplied with an additional factor $1/(1-\rho_d \gamma)$ to
match to the full wave function renormalization. By doing so, we have
\begin{equation}
\label{eq:deltaPol}
\delta^A_{\text{pol},L} = \frac{1}{1-\rho_d \gamma} \left(-0.962 + 0.009 + 0.004\right)~\text{meV}~= -1.605~\text{meV}\pm 0.066~\text{meV}~.
\end{equation}
$\mathcal{Z}_d$-improvement can also be applied to the evaluation of
the structure function, the result of which is shown in
Fig.~\ref{fig:SL}.  The uncertainty presented in
Eq.~\eqref{eq:deltaPol} is $\sim(\rho_{d}\gamma)^{3}$, where new
$\slashed{\pi}$EFT parameters at higher orders are expected to enter.

In Table \ref{tab:Shifts}, we show the $\slashed{\pi}$EFT results of
$\delta^A_{\text{pol},L}$ computed at LO, NLO, NNLO and with the NNLO
$\cZd$-improvement approach. We also shown the comparison with
Ref.~\cite{Hernandez:2019zcm} in the table, where the calculations
were done using two different expansion methods, named $\eta$-less
method and $\eta$-expansion method respectively.  Using the
$\eta$-less method, Ref.~\cite{Hernandez:2019zcm} evaluated
$\delta^A_{\text{pol},L}$ in the multipole expansion of the charge
operator and summing up multipole contributions to high orders. In the
$\eta$-expansion approach, $\delta^A_{\text{pol},L}$ in the same
non-relativistic and point-proton limit is equivalent to
$\delta^{A}_{{\rm pol},L} =
\delta^{(0)}_{D1}+\delta^{(1)}_{Z3}+\delta^{(2)}_{R2}+\delta^{(2)}_{Q}+\delta^{(2)}_{D1D3}$
with notations given in Ref.~\cite{Ji:2018ozm}. Using nuclear
potentials in the ZRA, $\eta$-less and $\eta$-expansion methods both
obtained $\delta^A_{\text{pol},L} = -1.590$
meV~\cite{Hernandez:2019zcm}, which is different by $0.8\%$ from this
work. Both $\slashed{\pi}$EFT and ZRA are based on the effective range
expansion. However, in ZRA, only the bound-state wave function
renormalization is range corrected as in $\cZd$. In the work by
Hernandez {\it et al.}, the intermediate scattering states in the
two-photon processes were treated as plane waves subtracted by the
bound ground state when using the ZRA potential. Using $\chi$EFT
potential, where final-state interactions are embedded in the
diagonalization of the nuclear Hamiltonian, $\delta^A_{\text{pol},L}$
obtained in $\eta$-less and $\eta$-expansion methods are respectively
$-1.588$ and $-1.590$ meV~\cite{Hernandez:2019zcm}. The results from
$\slashed{\pi}$EFT agree with both ZRA and $\chi$EFT calculations
within the expected uncertainty at NNLO.

The energy shift can be similarly evaluated in the relativistic limit
by using the relativistic kernel from Eq.~\eqref{eq:KL} in the sum
rule integration.  To obtain the $\eta$-expansion results of the
relativistic $\delta^{A}_{{\rm pol},L}$ in ZRA (or $\chi$EFT), we need
to add an additional relativistic correction $0.037$ meV from
Ref.~\cite{Friar2013} (or $0.030$ meV from Ref.~\cite{Ji:2018ozm}) to
the non-relativistic value. The resulting $\delta^{A}_{{\rm pol},L}$
is $-1.553$ meV in ZRA and $-1.560$ meV in $\chi$EFT. The relativistic
$\delta^{A}_{{\rm pol},L}$ calculated using $\eta$-less method (noted
by $\Delta_L$ in Ref.~\cite{Hernandez:2019zcm}) is $-1.562$ meV in
$\chi$EFT. We calculate the relativistic $\eta$-less
$\delta^{A}_{{\rm pol},L}$ in ZRA by using the analytic matrix element
from Ref.~\cite{Hernandez:2019zcm} and obtain $-1.564$ meV.  Results
with $\slashed{\pi}$EFT in relativistic limit and comparison are also
shown in Table~\ref{tab:Shifts}. It indicates that agreement within a
$~1\%$ discrepancy in $\delta^{A}_{{\rm pol},L}$ is also achieved in
the relativistic limit.

\section{Summary}
\label{sec:summary}
In this work, we have calculated the longitudinal structure function to
NNLO in $\slashed{\pi}$EFT. We have given analytic expressions for the squared
matrix element required to calculate the structure function. At NNLO,
only two parameters, {\it i.e.}, the deuteron binding momentum and
S-wave spin-triplet effective range, are required as experimental
input.  We furthermore included final and initial state interactions
consistently and showed explicitly that final state interactions are
strongly suppressed at higher orders. We separated out the
dipole contribution to quantify how much S-wave final state
interactions contribute to this process. The $\slashed{\pi}$EFT approach brings the
advantage that calculations can be done largely analytical in the
two-nucleon sector, and very few parameters are directly
related to two-nucleon scattering enter the calculation. At NNLO, we
expect for an accuracy of approximately 5\%.

We compared our results for the energy shift
$\delta_{{\rm pol},L}^{A}$ in $\mu$-d with recent calculations using
$\chi$EFT and ZRA nuclear potentials.  We found good
agreement with these calculations and confirm thereby our uncertainty
estimate.  We limited ourselves to NNLO because the deuteron
channel shape parameter, two-body current counterterm and S-D mixing operator must be
included at higher orders, making the analysis significantly more
complicated. However, while much more involved, such a higher order
calculation might also lead to interesting results, specifically if
the unknown counterterm could be used to describe universal
correlations between electromagnetic observables such as the deuteron
radius and the nuclear polarizability correction.

Another extension of this work is to calculate the TPE
contribution to the Lamb shift in the triton and Helium-3. 
Recent calculations show that the $\slashed{\pi}$EFT is well-suited to describe 
the electromagnetic properties of the three-nucleon system~\cite{Vanasse:2015fph,Vanasse:2017kgh}. In the 
three-nucleon system, $\slashed{\pi}$EFT loses some of its advantages. 
For example, the wave function has
to be calculated numerically, too. We also expect significantly slower
convergence and therefore larger uncertainties as the binding momentum
of the three-nucleon states is much larger than the one of the
deuteron. However, it is an important and interesting question by
itself how well such details of the three-nucleon state can be
described in $\slashed{\pi}$EFT.

\begin{acknowledgments}
  We thank Richard Hill for discussions in the early stages of this
  project and Sonia Bacca for comments on this manuscript. This work
  has been funded by the National Science Foundation under Grant No.
  PHY-1555030, by the Office of Nuclear Physics, U.S.  Department of
  Energy under Contract No. DE-AC05-00OR22725 and by the National
  Natural Science Foundation of China under Grant No. 11805078.
\end{acknowledgments}

\begin{appendix}
\section{Relevant loop integrals in the power-divergence subtraction
  scheme}
\subsection{Two-point loop integrals}
We define the two-point loop integrals that are required in the calculation 
up through NNLO as in Ref.~\cite{Kaplan:1998tg}, such that
\begin{eqnarray}
\label{eq:I2nPDS}
\cI_{2n}^{\rm PDS} (E)
&=& i \int \frac{\hbox{d}^4 q}{(2\pi)^4} q^{2n} S(q_0+E,\bs{q}) S(-q_0,-\bs{q}) \nn\\
&=& -\frac{m_N }{4\pi}p^{2n}\left(\mu + i p\right)~,
\end{eqnarray}
where $p = \sqrt{m_N E}$ is the relative momentum within the two
nucleon pair. For deuteron bound state we have $p=i\gamma$, with
$\gamma = \sqrt{m_N B_d}$ denoting the deuteron binding momentum.

\subsection{Loop integral $\cJ_0$}
\label{sec:J0}
In diagram (c) of Fig.~\ref{fig:MLO}, the three-point loop integral $\cJ_0$
needed in Eq.~\eqref{eq:Mc-short} is given as 
\begin{equation}
    \label{eq:J0appendix}
      \cJ_{2n} =\int \frac{\hbox{d}^4 l}{(2\pi)^4} l^{2n} iS(-B_d+l_0,\bs{l})
i S(-B_d+l_0+\omega, \bs{l}+{\bf q}) i S(-l_0,\bs{l})~,
\end{equation}
which has three nucleon propagators in the integrand. This integral 
can be solved in various ways. One way that relates it
to the calculation of a quantum mechanical matrix element of the
charge operator is to solve it in coordinate space.  We can reexpress
the diagram as an integral over to coordinate space wave
functions using Fourier transform of the two propagators
\begin{align}
  \label{eq:GreenFT1}
  \frac{1}{-B_d - \frac{k^2}{m_N} + i \epsilon} &= -m_N \int \hbox{d}^3 r 
  e^{i\bs{k}\cdot \bs{r}} \frac{ e^{ -\gamma r}}{4 \pi r}~,\nn\\
\frac{1}{E-\frac{(\bs{k}+\bs{q}/2)^2}{m_N} + i\epsilon }
& = - m_N \int \hbox{d}^3 r e^{i (\bs{k}+\bs{q}/2)\cdot \bs{r} } 
\frac{ e^{ i p r}}{4 \pi r}~,
\end{align}
where $E = p^2/m_N = \omega - B_d -q^2/(4m_N)$. For $\cJ_0$, placing the results 
of Eq.~\eqref{eq:GreenFT1} in Eq.~\eqref{eq:J0appendix} gives 
\begin{align}
  \label{eq:J0coord}
  \cJ_0 = - 8\pi \left(\frac{m_N}{4\pi}\right)^2\int \hbox{d} r e^{-\gamma
  r} e^{i p r} \frac{\sin(q r/2)}{q r}~.
\end{align}
Evaluating the integral in Eq.~\eqref{eq:J0coord}, we find
\begin{align}
  \label{eq:J0coord2}
  \mathcal{J}_0 = -\frac{m_N^2}{4\pi q} 
  \left[ 
  \cot^{-1}\left(\frac{m_N \omega -q^2/2}{q\gamma} \right) 
  + i \tanh^{-1} \left( \frac{q \sqrt{m_N \omega -\gamma^2 -q^2/4}}{m_N \omega}\right)
  \right]~,
\end{align}
where we define the range of $\cot^{-1}$ as $(0,\pi)$.

In the evaluation of the matrix elements in this work we encounter the
sum $\cJ_2 +\gamma^2 \cJ_0$. We can express this sum as
\begin{align}
  \label{eq:J2-gJ0-1}
  \cJ_2 +\gamma^2 \cJ_0  = m_N^2\int \frac{\hbox{d}^3 k}{(2\pi)^3}
  \frac{1}{m_N\omega-\gamma^2-q^2/4  -k^2+i\epsilon}~.
\end{align}
This integral can be solved using the PDS formula from KSW
\begin{align}
  \label{eq:J2-gJ0-pds}
  \left(\cJ_2 +\gamma^2 \cJ_0\right)_{\rm PDS}  =-\frac{
  m_N^2}{4\pi}\left(
\mu+i\sqrt{m_N\omega-\gamma^2-\frac{q^2}{4}}\right) = m_N  \cI_0 (E)~,
\end{align}
with $m_N E= m_N\omega-\gamma^2-q^2/4 = p^2$ due to energy conservation.
\subsection{Loop integral $\widetilde{\cJ}_2$}
\label{sec:J2}
At NLO, we encounter the loop diagram $\widetilde{\cJ}_2$ in the loop
that has NLO final state interactions. We define it as
\begin{equation}
  \label{eq:J2bar}
      \tilde{\cJ}_2=\int \frac{\hbox{d}^4 k}{(2\pi)^4} ({\bf
         k}+\frac{\bs{q}}{2})^2 iS(-B_d+k_0,\bs{k})
i S(-B_d+k_0+\omega, \bs{k}+\bs{q}) i S(-k_0,-\bs{k})~.
\end{equation}
Carrying out the countour integration gives
\begin{equation}
  \label{eq:J2bar-4}
  \tilde{\cJ}_2=\int \frac{\hbox{d}^3 k}{(2\pi)^3} ({\bf
    k}+\frac{\bs{q}}{2})^2
  i\left[
    -B_d-\frac{k^2}{m_N} + i\epsilon
  \right]^{-1}
  i
  \left[
    -B_d+\omega-\frac{q^2}{4m_N}-\frac{1}{m_N}(\bs{k}+\frac{\bs{q}}{2})^2+i\epsilon
  \right]^{-1}~.
\end{equation}
We can now evaluate the relevant sum that involves $\widetilde{\cJ}_2$
\begin{align}
  \nonumber
  \label{eq:J2bar-sum-1}
 \tilde{\cJ}_2 -p^2 \cJ_0 =&\int\frac{\hbox{d}^3 k}{(2\pi)^3}
  \left[({\bf k}+\frac{\bs{q}}{2})^2-\frac{p^2}{2}\right]
  i\left[-B_d-\frac{k^2}{m_N} + i\epsilon \right]^{-1}\\
  &\qquad\qquad \times i \left[
    -B_d+\omega-\frac{q^2}{4m_N}-\frac{1}{m_N}(\bs{k}+\frac{\bs{q}}{2})^2+i\epsilon
    \right]^{-1} \nn \\
   &=\frac{m_N^2}{4\pi}(\gamma-\mu)~,
\end{align}
where we used $p^2 = m_N (\omega -B_d) -q^2/4$ to rewrite the denominator 
in the last factor in the integrand of Eq.~\eqref{eq:J2bar-sum-1}. 
Similarly, one can show that
\begin{equation}
\label{eq:tildeJ2-2}
\tilde{\cJ}_2 +\gamma^2 \cJ_0 =m_N \cI_0(E)=-\frac{m_N^2}{4\pi}(\mu+i p)~.
\end{equation}

\end{appendix}
\bibliographystyle{apsrev4-1}
\bibliography{manuscript.bbl}
\end{document}